\documentclass{article}
\usepackage[utf8]{inputenc}

\usepackage{hyperref}

\title{End to End Software Engineering Research}

\author{
  Idan Amit\\
  Department of Computer Science\\
  Hebrew University of Jerusalem and Acumen Labs \\
  {\tt\small idan.amit@mail.huji.ac.il}\\
}
\begin{document}

\maketitle

\begin{abstract}
End to end learning is machine learning starting in raw data and predicting a desired concept, with all steps done automatically.
In software engineering context, we see it as starting from the source code and predicting process metrics.
This framework can be used for predicting defects, code quality, productivity and more.
End-to-end improves over features based machine learning by not requiring domain experts and being able to extract new knowledge.
We describe a dataset of 5M files from 15k projects constructed for this goal.
The dataset is constructed in a way that enables not only predicting concepts but also investigating their causes.
\end{abstract}

\section{Introduction}

Our long term goal is to investigate causal relations in software engineering.
In which length are files less prone to error?
Which code smells \cite{1999:RID:311424} should be removed in order to improve productivity \cite{amit2021follow}?

We present a dataset designed to investigate such questions.
The dataset is built for end to end learning, starting in the source code and ending in various process metrics \cite{Moser:2008:ARS:1414004.1414063,6606589}.
The dataset samples the source code every two months, starting in June 2021.
That enables to use co-change analysis \cite{Amit2021CCP, amit2021follow} in order to find patterns (e.g., code smells) that are not only predictive but also more likely to be causal. 
We further discuss the use for causality investigation in Section \ref{sec:Causality}.

Empirical software engineering needs such large datasets that can serve as a common benchmark.
Deep learning in software engineering suffers from replicability and reproducibility problems \cite{liu2021reproducibility}.
Given the dataset, replication research on the same dataset is easier.
Due to the dataset for different languages and dates, reproduction is also easier.

A large part of the research is based on metrics and features  designed by experts.
Designing a metric is time consuming and hence there are not many of them.
Even the best metric suite cannot capture all aspects of interest.
By using end to end learning we avoid this problem.

We are far from exploring the large number of possible supervised learning concepts.
While defect prediction is a very common use case, we can use the dataset to investigate quality, productivity, security and many other process metrics.

\begin {table}[h!]\centering
\caption{ \label{tab:dataset-size} Dataset Size }
\begin{tabular} { | l| c| c|  }
\hline
\textbf{Language} & \textbf{Projects} & \textbf{Files}\\
\hline
C & 1,121 & 614,671\\ \hline
C\# & 952 & 488,920\\ \hline
C++ & 1,609 & 571,962\\ \hline
Go & 845 & 337,664\\ \hline
Java & 2,259 & 1,763,882\\ \hline
JavaScript & 3,133 & 621,434\\ \hline
PHP & 1,607 & 729,014\\ \hline
Python & 3,592 & 560,965\\ \hline
All & 15,118 & 5,688,512\\ \hline
\end{tabular}
\end{table}

Table \ref{tab:dataset-size} presents the projects and files extracted in June 2021.
The extraction is based on the process described in \cite{Amit2021CCP}, using the infrastructure \cite{Amit2021General} and the specific code from \cite{Amit2021E2eseSrc}.
The process filters out tiny, not up-to-date,  redundant, and not software projects.

\section{Related Work}
\label{sec:related}

End to end learning is common in deep learning, due to its ability to extract representations.
See \cite{watson2021systematic} for a systematic review of 128 papers using deep learning for 23 software engineering tasks.

Datasets are common in software engineering today.
Some of them collect diverse data on projects, aiming to enable multiple uses \cite{gousios2012ghtorrent, Lenarduzzi_2019,Trautsch2017, Amit2021General}.

Others are dedicated to a specific task, and usually come with features and labels to support it.
Datasets of 150k JavaScript files \cite{10.1145/2914770.2837671} and 150k methods \cite{murali2018neural} were used for program generation.
More examples are datasets for vulnerability \cite{10.1145/3468264.3473122, Amit2021Sweets}, tangling commits \cite{herbold2021fine}, text summarization \cite{choshen2021comsum}, Truck Factor Developers Detachment (TFDD) \cite{avelino2019abandonment}, commit classification \cite{Amit2021CommitClassification, 10.1145/3127005.3127016}, code translation \cite{lu2021codexglue} and many more \cite{Sayyad-Shirabad+Menzies:2005, sharma2021survey}.

Another common type of datasets collections of bugs and defect prediction datasets \cite{just2014defects4j, 8730197, 8595167, Amit2021Sweets}.

We built this dataset in order to enable both general exploration and specific tasks.
We pre-computed many process metrics, allowing researchers to build supervised learning models for defect prediction, quality, vulnerability, productivity, etc.
We also provided the source code and commit messages.
This allows future definition of new process metrics and adapting the dataset for predicting it.
We also suggest new tasks that the dataset can be used for.

\section{Use Cases}
\label{sec:use-case}

\subsection{Defect Prediction}

Defect prediction is the application of machine learning in order to find future bugs \cite{5463279,  Hall:2012:SLR:2420627.2420790}.
Bug identification is important in order to reduce the cost of errors, focus testing effort, and decide on refactor targets. That led to hundreds of papers investigating this area.

Our dataset enables defect prediction by supplying bugs found per file in the detection period.
While being the common way to model defect prediction, we believe that this approach suffers from many drawbacks and suggest predicting the more robust Corrective Commit Probability (CCP)\cite{Amit2021CCP} instead.

Typically defect prediction requires a single bug to consider a file as buggy.
That is problematic in both directions.
A single bug might be accidental.
On the other hand, we do not differ between the severity of a single bug and five bugs.
This problem is solved sometimes by predicting the number of bugs.
However, this way we have equivalence between a file that was modified once to fix a bug and a file modified 20 times yet fixing a bug once.

A simple solution is to normalize, and that is what the corrective commit probability \cite{Amit2021CCP} does, normalize the bug fixes in the number of commits.
In defect prediction we deal with a small number of commits.
For larger numbers, as when estimating an entire project, maximum likelihood estimation is added to compensate for the bug-fix language model's mistakes.
The maximum likelihood adaptation is linear and monotonically increasing and therefore does not change the results in defect prediction.

Many tasks benefit from large datasets \cite{prenner2021making} and so does defect prediction \cite{Alshayeb_Alshammari_2021}.
We provide a large dataset for defect prediction.
More than that, since CCP is available for many more files, one can build a model for past CCP and use it downstream as part of future defect prediction. 

\subsection{Process metrics}

Defect prediction is a single narrow concept that can be predicted on code.
There are plenty of other interesting important concepts: run time performance issues, productivity, security vulnerabilities and more.

We provide such process metrics based on meta data - analysis of commit messages, files in commit, developer's workflow, etc. 

Process metrics have been claimed to show where effort is being invested and having less stagnation \cite{Moser:2008:ARS:1414004.1414063,6606589}.
Moreover, process metrics are conditionally independent \cite{Blum:1998:CLU:279943.279962, 10.1007/BFb0026666} from code metrics.
In particular, that makes them conditionally independent from the programming language, file length, code smells, etc.

\subsection{Causality}\label{sec:Causality}

We are not interested in only predicting code quality.
We would like to find ways to improve code quality.
In order to know how to improve, we should identify causal relations.

Our dataset enables the use of many methodologies in order to find and validate causal relations.
Using data analysis and machine learning, one can find patterns related to high quality.
Everyone is aware that ``Correlation does not imply causation''.
However, causation is translated into correlation (possibly with complex conditioning), hence correlative relations are a good start when looking for causation.

Conditioning is a good way to factor-out confounding variables.
One can use twin experiments \cite{vandenberg1966contributions} by comparing the behavior of the same developer in two different contexts (e.g., different projects) \cite{Amit2021CCP, amit2021follow, Amit2021Analysis}.
That allows to factor-out the developer skill, style, etc. without even knowing them.

Co-change analysis \cite{Amit2021CCP, amit2021follow, Amit2021Analysis} allows to use different behaviours of the same entity changes over time.
Given a causal relation, a change in the cause should lead to a change in the effect (possibly conditioned).
If there is no such relation, such co-change might be due to other reasons (e.g., a common cause) but otherwise less common.
One can use the cause improvement as a classifier and effect improvement as a concept and use the precision lift to identify likely causes.
The use of control variables can remove the threat of common causes.

If a fixed set of variables is used for prediction, it is easy to extend the method.
Assuming the Markov property \cite{frydenberg1990chain}, we can focus in the current and previous states.
One can use both the current set of variables and the difference between the current variables and the variable in the previous timestamp.

If we have a predictive model with perfect accuracy, we are assured that there is no other causal variable (in this data set), since its existence will prevent perfect prediction.
If we have a minimal predictive model, each variable change leads to a change in prediction in at least one flow (if not, it could have been removed hence the model is not minimal).
Therefore a perfect minimal predictive model means there are no other causes and for each variable there is a case where changing it changes the prediction.

However, many machine learning modeling problems are NP-complete \cite{10.1145/800157.805047, chickering1994learning, chickering1996learning, 10.5555/2969735.2969792, laurent1976constructing}.
In some cases, even weak learning is not feasible \cite{schapire1990strength, kearns1994cryptographic}.
Therefore, we cannot expect to find \emph{a minimal model}.
However, finding \emph{a model} can shed light on unknown relations.

When observing changes in nature, there are few common uncertainties.
We cannot know why a developer chooses to perform a change (e.g., preference to low hanging fruits or acting only in crisis).
We are not sure if the change is a local clean change or a total rewrite.
Even a clean change might be later interfered by more changes.
The classical test-bed for causality is the controlled experiment, aiming to deal with these changes.

Choosing cases for experiments is a harder than expected problem in software engineering.
A researcher might build toy problems yet they usually do not resemble how code is developed and maintained.
Finding suitable cases manually is hard since usually there are not many of them.

Our dataset enables intervention experiments \cite{amit2021follow}.
Consider the removal of code smell as an example.
One can look for source code having the smell.
One can also assure that the context is valid for change analysis (e.g., long enough prior history, no major refactoring before the current date).
Given the potential intervention points, the researcher can choose the cases to act on and a control group by a desired policy. 
Then, perform the intervention and investigate its result.

\subsection{Program Repair}

In program repair, the goal is to automatically generate a modification that fixes a bug \cite{nguyen2013semfix, le2016history,goues2019automated}.
Similarly, the goal of automatic refactoring is to generate a refactor modification that keeps the program functionality yet improves its structure \cite{ge2012reconciling}.

Samples of such cases are important in order to build relevant models and evaluate them.
We identify these cases in files that were modified only once in the modification period, attributing the change to a single commit.
We also require the commit to modify only one source code file (optionally with one test file), in order to be sure that the modification to the file is self-contained.

This dataset also has a surprising educational value.
Developers rarely get feedback on bugs that they miss or beneficial refactors not applied.
This dataset can be used as a question/answer training.
It is likely that not all answers (the modification done to the prior version), should be accepted as canonical answers.
Filtering can be done by selecting only answers improving a desired metric.
Another solution is to use the wisdom of the crowd and use the developers feedback to identify these answers.

\subsection{Code Similarity}

Comparison of software has a natural threat of comparing apples and oranges.
Hence when trying to compare aspects like programming language might not compare similar projects \cite{876288,7194625, Ray:2014:LSS:2635868.2635922, 10.1145/3340571, 6032456, 7476675}.
Also, projects in different domains have a different tendency to be error prone, even when in the same language \cite{Amit2021CCP}.

One approach to have similarity of programs is to use different solutions to the same task \cite{puri2021codenet, 8812669, mou2015convolutional}.
One can use these dataset to construct a code similarity metric \cite{ye2021misim}.
That enables comparing similar programs in different settings (e.g., programming language), focusing only on the differing aspect.

However, the main difficulty is the definition of similar code.
This is not well defined and while there are easy cases of surely similar or not similar, the gray area is large.
We suggest using different versions of the same file as positive examples of similar code.
This suggestion relies on the fact that the developer considers them to be similar, by putting them in the same file.
In principle, any pair of different files can serve as negatives.
However, this will lead to a positive rate of $O(\frac{1}{n})$ which is both very imbalanced \cite{van2007experimental, krawczyk2016learning, 10.1145/3338501.3357374} and rather easy since most pairs will probably be different in many aspects.
Instead, we consider only files from the same project and the same directory, increasing the challenge by comparing close files.

As we collect versions in more dates, we can require a higher bar.
The identifiers are informative in code similarity \cite{ahmed2021multilingual}, and by using versions of the same file we reduce their benefit and force the model to use different information.
We can do it by comparing a version of a file to two other versions of it and require the model to find the  \emph{closer} version.
Hence the October version of the file should be closer to the December version than to the June version.
The concept of closer can be computed by the date difference between the versions or the number of commits separating them.
The similarity dataset contains all the files in the projects, hence can be used for text similarity (e.g., in English) and image similarity in the same method.

\subsection{Program Difficulty}

Program difficulty is another aspect that should be considered when comparing software.
Once again, there is no common way to measure difficulty.
Sometimes a manual definition of task difficulty can be used to label the difficulty to its solutions \cite{puri2021codenet}.
We approach the problem by using some labeling functions \cite{NIPS2016_6523, archimedes} differing difficult and easy files.
The first labeling function that we use compares two files  written by only one same developer in the same project.
By doing that we factor out the developer skill and the problem domain.
We compare error prone files to a risk reduce file, measured by CCP.
This labeling function operationalizes the definition ``It is more likely to make mistakes in harder tasks''.

The second labeling function is inspired by the habit of ``Good first issues'' \cite{huang2021characterizing}.
It is common to let new developers in a project to work on easier tasks, helping them to get familiar with the project.
Again, we compare files owned by the same developer in the same project, one written in the developers first 6 months and one in the next 6 months.
We do not use later files in order to avoid the threat of significant skill improvement.
This labeling function operationalizes the definition ``Beginners tend to work on easier tasks''.

The last labeling function uses the amount of work invested in the file.
We use files written in a single commit, and use the gross work duration.
The gross time is the time from the developer's previous commit, given it was in the same day \cite{Amit2021CCP}.
This labeling function operationalizes the definition ``Harder tasks take longer''.

Each labeling function captures a different aspect of difficulty and uses a different dataset.
One can build models upon them and use methods that can aggregate them in order to have a combined difficulty measure \cite{NIPS2016_6523, archimedes}.

\subsection{Other Use Cases}

We provide the relevant commit messages in order to enable the use of new linguistic models for computing new process metrics.
The commit messages themselves can be used for text summarization \cite{choshen2021comsum}.

The project, and the topics that the contributors attach to them, can be used for topic identification.
The flow of developer activity (e.g., commit dates, modified files), can be used to identify a change in behaviour \cite{amit2020identifying}.

The dataset can be used to answer many descriptive statistics questions of interest.
One can find the distribution of LOC per function, the distribution of the number of parameters for function, etc.

Last but not least, the source code can be used to train language models.
These can be used for many downstream tasks from prediction to code generation completion \cite{hayati2018retrieval, chen2021evaluating}.
More uses are  training for code search \cite{gu2018deep}, and automatic documentation \cite{miceli2017parallel}.

\section{Threats to Validity}
\label{sec:threats}

We used language models \cite{Amit2021CommitClassification, Amit2021CCP, 10.1145/3345629.3345631} in order to analyze the commit messages and extract process metrics as bug fixes and refactors.
These models have high performance (e.g., the corrective model is close to human level) yet they are not perfect.
We use a maximum likelihood estimation method for classifiers in order to find the positive rate given the hit rate \cite{Amit2021CommitClassification} yet the result is still not perfect.

Though that dataset is stale, we would like to alert over changes that might influence it.
The data keeps evolving and therefore subject to concept drift \cite{gama2014survey}.
Hence reporting on the recent or all time datasets might be changed.
We hope that the language models will keep improving, hence existing commit messages might be classified differently in the future (while we keep the stale version).

We provide two kinds of splits to train, validation and test sets.
The larger split is done by repositories and each repository split is further split by files.
Note that the Independent and Identical Distribution assumption is not likely to hold when splitting by repositories since that might be very different.
We provide this split in order to investigate domain adaptation and cross project prediction, a severe problem in defect prediction \cite{zimmermann2009cross}.

The definition of the concept in supervised learning is always of high importance yet here it might be critical.
Consider the use case of defect prediction, a well explored area.
The concept seems trivial: "Did a defect appear in the file?".
However, defects are marked only after they are fixed.
Kim and Whithead reported that the median time to fix a bug is about 200 days \cite{Kim:2006:LDT:1137983.1138027}.

Hence, when using bug detection as a concept, a critical question is the length of the detection period.
We used two months as the default period, and provided the data and code to use a different period.

A more subtle important decision, which is easy to miss, is the definition of negatives, files without bugs.
It seems natural to choose all files without bugs as negatives but this leads to two difficulties.
First, the file might have a bug yet the file was not used (and possibly part of a project entirely not used).
Another problem is the creation of an imbalanced dataset \cite{van2007experimental, krawczyk2016learning, 10.1145/3338501.3357374}.
The number of files changed during a two months period might be lower than 10\% of the project files.
Class imbalance causes classifiers to favor the majority class and make metrics like accuracy misleading.

A suitable definition of the target concept comes from the desired use case and therefore there is no universal solution.
We believe that for most use cases, choosing only files modified during the period as negative is a good option.

Another threat is the use of the source code as input.
One approach can be to remove all comments in prepossessing since comments are irrelevant to the program execution.
However, cases like `TODO' comments are informative \cite{10.1145/3345629.3345631} yet not causal regarding the existence of bugs.
Again, this decision should be taken with respect to a use case. 

There are many powerful natural language models hence it is tempting to use them as is for source code.
One should note that source code contains many entities (variables, functions) whose name is not a word.
When working on NLP it is common to represent rare tokens as `unknown token'.
Whoever, if applied to code, will lose the important information that some statements refer to the same entity.
In many cases the entity name is a list of words, concatenated in underscore (e.g., `customers\_dictionary', `list\_of\_orders').
They provide valuable information on the entity type and relations among entities that might help if represented.

We would like to know how representative our dataset is.
We do that by choosing some organizations and checking how many of their open source projects appear in our dataset (the full one \cite{Amit2021CCP, Amit2021General} not just the currently exported), presented in Table \ref{tab:projects_representation}.

\begin {table}[h!]\centering
\caption{ \label{tab:projects_representation} Projects Representation }
\begin{tabular}{|l|l|l|l|}
\hline
\textbf{Organization} & \textbf{Projects} & \textbf{BigQurey} & \textbf{Dataset}  \\ \hline
Apache & 2.3k & 420 & 94  \\ \hline
Google & 2.1k & 1789 &  248  \\ \hline
\href{https://github.com/wikimedia}{Wikimedia} & 2.4k & 480 &  96  \\ \hline
GNOME & 381 & 401 &  78  \\ \hline
Tensorflow & 101 & 93 &  48  \\ \hline
\end{tabular}
\end{table}

Hence the GitHub BigQuery schema recall seems to be in the range of 20\%-90\% (GNOME probably deleted projects that still exist in the schema).
The requirements of the dataset filter (e.g., at least 50 commits in 2020) filters out 50\%-80\% of the projects of these rather active organisations.
Note that only 1\% of the projects in BigQuery have at least 50 commits in 2020.

While we try to capture the state of the project in a given time, noise exists.
53 commits out of 59 million had a timestamp higher than the one in which they were fetched.
Hence, this specific problem is negligible yet other timing problems probably exist. 
Developers might commit at a given time, send their work to review and have it accepted after a long time.
BiqQuery updates the GitHub schema once a week, which might also lead to mismatches.
In cases where the precise state is needed we recommend cloning the project and to examine the project at that time.

\section{Future Work}
\label{sec:future}

The programming languages in the dataset represent the 20 most common source code extensions.
More languages exist and might be of interest.

This dataset is still evolving.
We started publishing data and we plan to further extend it.
We will keep publishing future versions.

We are interested in ideas, use cases and extensions of the datasets.
If you have such, please contact us.
If you have datasets or tools that can be used for enhancement, we will be interested too.

We provide all the source code used to generate the dataset.
This way any researcher can construct datasets fitting for specific needs.
That makes decisions taken (e.g., a gap of two months between source code versions), more flexible and less important.
We hope that the dataset will lead to plenty of future work in the presented use cases and many new ones.

\section{Conclusion}
\label{sect:conclusion}

\begin{itemize}
  \item We present a large and diverse dataset that enables end to end learning in software engineering.
  \item The dataset extracts source code and process metrics every two month to enable investigating causality.
  \item We provide labeling functions and dataset to investigate code similarity.
  \item We provide labeling functions and dataset to investigate program difficulty.
\end{itemize}

\section*{Supplementary Materials}

The code used to produce the dataset is in \cite{Amit2021E2eseSrc} and so are links to the repositories containing the datasets.

\bibliographystyle{abbrv}
\bibliography{bibtex.bib}

\end{document}